\documentclass[superscriptaddress,twocolumn]{revtex4}
\usepackage{graphicx}
\usepackage{color}

\renewcommand{\vec}[1]{\mbox{\boldmath $#1$}}

\begin{document}

\title{
Role of deformation in odd-even staggering in reaction cross 
sections for $^{30,31,32}$Ne and $^{36,37,38}$Mg isotopes}

\author{Y. Urata}
\affiliation{ 
Department of Physics, Tohoku University, Sendai, 980-8578,  Japan} 

\author{K. Hagino}
\affiliation{ 
Department of Physics, Tohoku University, Sendai, 980-8578,  Japan} 
\affiliation{Research Center for Electron Photon Science, Tohoku University, 1-2-1 Mikamine, Sendai 982-0826, Japan}

\author{H. Sagawa}
\affiliation{
RIKEN Nishina Center, Wako 351-0198, Japan}
\affiliation{
Center for Mathematics and Physics,  University of Aizu, 
Aizu-Wakamatsu, Fukushima 965-8560,  Japan}


\begin{abstract}
We discuss the role of pairing anti-halo effect in the observed odd-even 
staggering in reaction cross sections for 
$^{30,31,32}$Ne and $^{36,37,38}$Mg isotopes by taking into account 
the ground state deformation of these nuclei. 
To this end, we construct the ground state density 
for the $^{30,31}$Ne and $^{36,37}$Mg nuclei based on 
a deformed Woods-Saxon potential, while for the $^{32}$Ne and $^{38}$Mg
nuclei we also take into account the pairing correlation 
using the Hartree-Fock-Bogoliubov 
method. 
We demonstrate that, when the one-neutron separation energy is small for 
the odd-mass nuclei, a significant odd-even staggering 
still appears even with finite deformation, although 
the degree of staggering is somewhat 
reduced compared to the spherical case. This implies that the 
pairing anti-halo effect in general 
plays an important role in generating 
the odd-even staggering in reaction cross sections 
for weakly bound nuclei.  
\end{abstract}

\maketitle

\section{Introduction}

The halo structure is one of the most important 
phenomena in neutron-rich nuclei\cite{TSK13,HTS13}. 
This phenomenon is characterized by a spatially 
extended density distribution originated from weakly bound 
valence neutron(s). This was first discovered by 
Tanihata {\it et al.}, who observed considerably large 
interaction cross sections for $^{11}$Li, $^{11}$Be, and $^{14}$Be 
\cite{Tanihata85,Tanihata88}. 
Subsequently, a narrow momentum distribution was also discovered 
\cite{Kobayashi88} for a weakly bound nucleus, $^{11}$Li, 
establishing the concept of the halo structure. 
The heaviest halo nucleus discovered so far is $^{37}$Mg 
\cite{Kobayashi14,Takechi14}. 

For weakly bound nuclei with two valence neutrons, the pairing 
correlation between the valence neutrons may quench the halo 
structure \cite{BDP00}. That is, the pairing correlation alters the 
asymptotic behavior of the wave function for the valence neutrons, 
reducing the divergence feature of nuclear radii for $s$ and $p$ waves 
at zero separation energy \cite{RJM92,Jensen04,S92,SH15}. 
This effect is referred to as the pairing anti-halo effect, which can also 
be viewed as a generation of a 
spatially localized 
wave packet of quasi-particles 
originated from a 
coherent scattering of the valence neutrons to the continuum spectrum 
caused by the pairing interaction \cite{HS17}. 

In the previous publications, we have argued that 
the pairing anti-halo effect plays an important role in 
the odd-even staggering 
observed in interaction cross sections \cite{HS11,HS12,HS12-2}. 
That is, the experimental data have often shown a large odd-even 
staggering in interaction and reaction cross sections, in which 
cross sections for odd-mass nuclei are systematically larger than 
those for the neighboring even-mass nuclei \cite{Takechi14,Takechi12}. 
Using the Hartree-Fock-Bogoliubov (HFB) method with spherical symmetry, 
we have shown that the observed odd-even staggering can be largely 
accounted for in terms of the pairing anti-halo effect 
(see also Refs. \cite{Sasabe13,MY14}). 

In this paper, we extend our previous analyses by taking into account 
the ground state deformation of weakly bound nuclei. To this end, we 
study the odd-even staggering in the 
$^{30,31,32}$Ne and $^{36,37,38}$Mg isotopes, for which the 
$^{31}$Ne and $^{37}$Mg nuclei have been suggested to have a deformed halo 
structure with $p$ wave 
\cite{Kobayashi14,
Takechi14,Takechi12,Nakamura09,Nakamura14,Hamamoto10,Urata11,Urata12,
Minomo11,Minomo12,Minomo12-2,Watanabe14}. 

There are two 
possible effects of nuclear deformation on the 
odd-even staggering. Firstly, several angular momentum components are 
mixed in a deformed single-particle wave function for a valence 
neutron, reducing the $s$ and 
$p$ wave components in the wave function. This will reduce the 
radius of the $^{31}$Ne and $^{37}$Mg nuclei, 
somewhat quenching 
the odd-even staggering in the interaction and reaction cross sections. 
Secondly, the deformation may 
change the level density around 
the Fermi level, which would result in either an enhancement or a 
decrease  of the pairing 
correlation, depending on the position of the Fermi surface. 
This would eventually influence the magnitude of 
the pairing anti-halo effect,
thus the cross sections for $^{32}$Ne and $^{38}$Mg. 
The primary aim of this paper is to investigate how these two effects 
interplay with each other in actual cases and how the conclusion obtained in 
our previous analyses based on spherical symmetry is altered if the 
deformation is explicitly taken into account. 

The paper is organized as follows. In Sec. II, 
we briefly summarize the theoretical 
frameworks. 
In our calculations, we first generate the deformed ground state density using 
the HFB method with Woods-Saxon potentials, which is then used as 
an input to the Glauber theory in order 
to compute reactions cross sections. In Sec. III, we apply these 
frameworks to reaction cross sections for the 
$^{30,31,32}$Ne and $^{36,37,38}$Mg nuclei and discuss the role of deformation 
in the odd-even staggering in the reaction cross sections. 
We then summarize the paper in Sec. IV. 

\section{Theoretical Frameworks}

\subsection{Deformed density} 

We analyze the reaction cross sections for 
the $^{30,31,32}$Ne and $^{36,37,38}$Mg nuclei. Symbolically, 
we denote the three 
isotopes in each element 
as $A, A+1$, and $A+2$ systems, respectively. 
Our first task is to construct the ground state density of each nucleus 
by taking into account the deformation. 
For simplicity, we ignore the pairing correlation in the $A$ and $A+1$ 
systems, and construct the density distribution by putting the nucleons into 
the lowest $A$ 
and $A+1$ single-particle orbits in a deformed Woods-Saxon potential, 
respectively (we have confirmed that the reaction cross section for 
the $^{30}$Ne and $^{36}$Mg nuclei 
does not significantly change even if the pairing 
correlation is taken into account). 

For the $A+2$ systems, on the other hand, we take into 
account the pairing correlation with the HFB method. 
In the coordinate space representation,  
the HFB equations read \cite{Doba96,Doba84,Bulgac}
\begin{eqnarray}
 \left( \begin{array}{cc}
  \hat{h}-\lambda   &   \Delta(\vec{r})  \\
   \Delta(\vec{r})  &  -\hat{h}+\lambda  \end{array} 
   \right)
    \left( \begin{array}{c}
  U_i(\vec{r}) \\
 V_i(\vec{r})    \end{array} 
   \right)
 = E_i 
   \left( \begin{array}{c}
  U_i(\vec{r}) \\
  V_i(\vec{r})    \end{array} 
   \right),
\label{HFB}
\end{eqnarray}
where 
$\Delta(\vec{r})$ is the pairing potential, and $\lambda$ and 
$E_i$ are the chemical potential and a quasi-particle energy, respectively. 
$\hat{h}$ is a mean-field Hamiltonian given by 
\begin{equation}
\hat{h}=-\frac{\hbar^2}{2m}\nabla^2 + V(\vec{r}), 
\end{equation}
where $V(\vec{r})$ is a mean-field potential and $m$ is 
the nucleon mass. 
Here, we have assumed that the
nucleon-nucleon interaction is zero range, so that
the mean-field and the pairing potentials are both local. 
In this framework, the 
density distribution is given by 
\begin{equation}
\rho(\vec{r})=\sum_i|V_i(\vec{r})|^2. 
\end{equation}
The chemical potential $\lambda$ is determined so that the 
average particle number is $A+2$, that is, 
\begin{equation}
\int d\vec{r}\,\rho(\vec{r})=A+2. 
\end{equation}

For simplicity, we use a deformed Woods-Saxon potential for 
$V(\vec{r})$ and $\Delta(\vec{r})$ given by, 
\begin{eqnarray}
V(\vec{r})&=&V_0\left(f(r)-R_0\beta_2\,\frac{df(r)}{dr}\,Y_{20}(\theta_{rd})\right) 
\label{sppot}
\nonumber \\
&& 
+V_{\rm ls}
\,\frac{1}{r}
\frac{df(r)}{dr}(\vec{l}\cdot\vec{s}), \\
\Delta(\vec{r})&=&\Delta_0\left(f(r)-R_0\beta_2\,\frac{df(r)}{dr}\,
Y_{20}(\theta_{rd})\right) 
\label{pairpot}
\end{eqnarray}
with 
\begin{equation}
f(r)=\frac{1}{1+\exp[(r-R_0)/a]}. 
\end{equation}
Here, 
we have assumed axially symmetric quadrupole deformation with 
the deformation parameter of $\beta_2$, and denoted 
the angle between the vector $\vec{r}$ and 
the symmetry axis as $\theta_{rd}$. 
In the single-particle potential, 
we take into account only the spherical part of the spin-orbit 
potential. 
For the protons, we also add the spherical Coulomb interaction 
to the mean field potential, Eq. (\ref{sppot}), with the radius 
of $R_0$. 

We use the same values for the parameters for the single-particle 
potential as those listed in 
Table I in Ref.\cite{Shoji09}, except for the depth parameter $V_0$ 
for the configuration for the valence orbit, for which we adjust 
the value of $V_0$ so that the neutron separation energy for the $A+1$ 
nuclei is reproduced. For simplicity, we use the same value for the 
deformation parameter for all the three 
isotopes, $A, A+1$, and $A+2$. 
For the strength $\Delta_0$ for the neutron pairing potential, 
(\ref{pairpot}), we determine it according to 
\begin{equation}
\bar{\Delta} \equiv \frac{\int d\vec{r}\,\Delta(\vec{r})f(r)}
{\int d\vec{r}\,f(r)},
\label{avdelta}
\end{equation}
with the average pairing gap of $\bar{\Delta}=12/\sqrt{A+2}$ MeV. 
For the protons, we ignore the pairing correlation as they are 
deeply bound in the nuclei considered in this paper, and thus 
the effect 
of pairing correlation on the nuclear radius is 
expected to be small. 

We solve the HFB equations, (\ref{HFB}), by expanding the upper 
and the lower components of the quasi-particle wave functions as, 
\begin{eqnarray}
U_K(\vec{r})&=& \sum_{n,l,j}u_{nlj}\,\psi_{nljK}(\vec{r}), 
\label{HFB-U}
\\
V_K(\vec{r})&=& \sum_{n,l,j}v_{nlj}\,\psi_{nljK}(\vec{r}), 
\label{HFB-V}
\end{eqnarray}
where $\{\psi_{nljK}(\vec{r})\}$ are eigen-functions of the 
single-particle Hamiltonian $\hat{h}$ when the deformation 
parameter 
$\beta_2$ is set to zero. This wave function is characterized by the radial 
quantum number $n$, the orbital angular momentum $l$, the total 
single-particle angular momentum $j$, and its projection onto the symmetry 
axis, $K$. 
Notice that $K$ is conserved in the quasi-particle wave functions, since 
we assume the axial symmetry. 
In Eqs. (\ref{HFB-U}) and (\ref{HFB-V}), we include the continuum states 
up to 30 MeV above the Fermi energy, $\lambda$, by discretizing them with 
the box boundary condition at $R_{\rm box}=60$ fm. 

\subsection{Reaction cross sections}

We next consider collision of a deformed projectile nucleus 
with a spherical target nucleus, 
and compute the reaction cross sections, $\sigma_R$. 
To this end, we employ the Glauber theory, which is based on the eikonal 
approximation and the adiabatic 
approximation to the nucleonic motions \cite{Glauber,BD04}. 
In order to calculate reaction cross sections, we also apply the 
adiabatic approximation to the rotational motion of 
a deformed nuclei. 
That is, we first fix the orientation angle of the deformed nucleus 
and then take an average of the resultant cross section 
over all the orientation angles \cite{CT99,HT12}. The reaction cross 
sections are thus expressed as 
\begin{equation}
\sigma_R=\frac{1}{4\pi}\int d\vec{\Omega}\,\sigma_R(\vec{\Omega}),
\end{equation}
where $\vec{\Omega}$ is the angle of the symmetric axis of the deformed 
nucleus in the laboratory frame, and 
$\sigma_R(\vec{\Omega})$ is the reaction cross section for fixed 
$\vec{\Omega}$. 

In the Glauber theory, the reaction cross section is 
computed as \cite{OKYS01,OYS92,HSAK07,HSCB10,HINS12,HHES16},
\begin{equation}
\sigma_R(\vec{\Omega})=\int d\vec{b}\,
\left(1-\left|e^{i\chi(\vec{b};\vec{\Omega})}\right|^2\right), 
\end{equation}
where $\vec{b}$ is the impact parameter and the phase shift function $\chi$ is 
given by 
\begin{eqnarray}
&&i\chi(\vec{b};\vec{\Omega})=
-\int d\vec{r}\,\rho_P(\vec{r};\vec{\Omega})\nonumber \\
&&\times\left[
1-\exp\left(-\int d\vec{r}'\rho_T(\vec{r}')\,
\Gamma_{NN}(\vec{s}-\vec{s}'+\vec{b})\right)\right]. \nonumber \\
\label{Glauber}
\end{eqnarray}
Here, 
$\vec{s}$ and $\vec{s}'$ are the transverse component 
(that is, the component which is parallel to $\vec{b}$) 
of $\vec{r}$ and $\vec{r}'$, respectively. 
$\Gamma_{NN}$ is the profile function for the $NN$ scattering, 
which we assume to be 
\cite{OKYS01,OYS92,HSAK07,HSCB10,HINS12,HHES16} 
\begin{equation}
\Gamma_{NN}(\vec{b})=\frac{1-i\alpha}{4\pi\beta}\,\sigma_{NN}\,
\exp\left(-\frac{b^2}{2\beta}\right),
\end{equation}
where $\sigma_{NN}$ is the total $NN$ cross section. 
In Eq. (\ref{Glauber}), 
$\rho_P$ and $\rho_T$ are the density distribution for 
the projectile and the 
target nuclei, respectively. 
We assume that the target density is spherical, 
$\rho_T(\vec{r})=\rho_T(r)$, while the projectile density 
has axial symmetry, that is, 
\begin{equation}
\rho_P(\vec{r};\vec{\Omega})=\rho_P(r,\theta_{rd}). 
\end{equation}
The projectile density can be expanded into multipoles as 
\begin{eqnarray}
\rho_P(r,\theta_{rd})&=&\sum_\lambda\rho^{(P)}_\lambda(r)Y_{\lambda 0}(\theta_{rd}), \\
&=&\sum_{\lambda,\mu}\sqrt{\frac{4\pi}{2\lambda+1}}\, 
\rho^{(P)}_\lambda(r)Y_{\lambda \mu}(\hat{\vec{r}})Y^*_{\lambda \mu}(\vec{\Omega}).
\end{eqnarray}
Notice that the phase shift function given by Eq. (\ref{Glauber}) takes 
into account the effect beyond the optical limit approximation following 
the prescription proposed in Ref. \cite{AIS00}. 
We evaluate it using the Fourier transform method \cite{HS12,BS95}. 

In this paper, we analyze the experimental data at incident energy $E$ = 240 
MeV/nucleon with $^{12}$C target \cite{Takechi12,Takechi14}. 
We use the same density for $^{12}$C as that 
given in Ref. \cite{OYS92}, while 
we use the same parameters given in Ref. \cite{Ibrahim08} 
for the profile function, $\Gamma_{NN}$. 

\section{Odd-Even staggering in reaction cross sections}

\subsection{$^{30,31,32}$Ne isotopes}

Let us now numerically evaluate the reaction cross section for deformed 
nuclei and discuss the role of deformation in the odd-even staggering in 
the reaction cross sections. We first consider the $^{30,31,32}$Ne isotopes, 
for which the odd-even staggering has been investigated assuming spherical 
symmetry \cite{HS11}. 

In Ref. \cite{Urata12}, we have shown that the measured 
reaction cross section for the $^{31}$Ne nucleus can be 
reproduced both with the particle-rotor model and with the Nilsson 
model with a deformed Woods-Saxon potential when the quadrupole 
deformation parameter is in the range of $0.17 \leq \beta_2 \leq 0.33$. 
In this case, the valence neutron in $^{31}$Ne 
occupies the [330 1/2] ($K^\pi=1/2^-$) 
orbit, which is connected to the $1f_{7/2}$ level in the 
spherical limit \cite{Hamamoto10}. 
In this paper, we therefore choose $\beta_2=0.3$. 
As has been demonstrated in Ref. \cite{Urata12}, the dependence of 
reaction cross sections on the deformation parameter is weak once 
the configuration of the valence orbit is fixed. 

\begin{figure} 
\includegraphics[scale=0.6,clip]{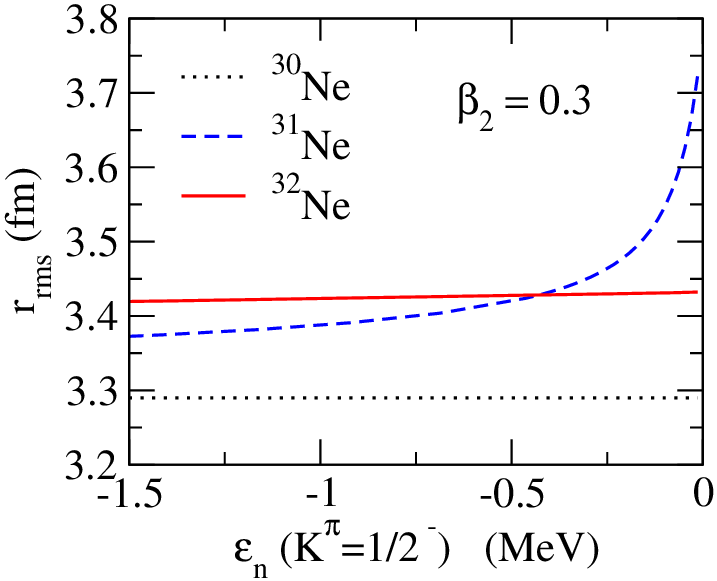}
\caption{The root-mean-square radius for the $^{30,31,32}$Ne isotopes as 
a function of the one neutron separation energy of the $^{31}$Ne nucleus. 
The dotted, the dashed, and the solid lines indicate the radius of the 
$^{30}$Ne, $^{31}$Ne, and $^{32}$Ne nuclei, respectively. 
The pairing correlation is taken into account only in the $^{32}$Ne 
nucleus using the Hartree-Fock-Bogoliubov method. 
The deformation 
parameter is set to be $\beta_2=0.3$ for all the three nuclei so that 
the valence neutron in $^{31}$Ne occupies the $K^\pi=1/2^-$ orbit. 
The figure is obtained by varying the depth parameter in the single-particle 
potential for the $K^\pi=1/2^-$ configuration. 
}
\end{figure}

Figure 1 shows the root mean square radii so obtained 
for the $^{30,31,32}$Ne nuclei 
as a function of the energy of the valence orbit for the $^{31}$Ne 
nucleus, $\epsilon_n$. 
To draw this figure, we vary the depth parameter, $V_0$, in the 
Woods-Saxon potential for the $K^\pi=1/2^-$ configuration. 
The dotted, the dashed, and the solid lines show the radius for the 
$^{30}$Ne, $^{31}$Ne, and $^{32}$Ne nuclei, respectively. 
One can see that the radius of the $^{31}$Ne increases rapidly 
as the one neutron separation energy, $S_n=-\epsilon_n$, 
approaches to zero, due to the $p$ wave 
component in the wave function for the valence neutron. On the other hand, 
the radius of the $^{32}$Ne nucleus varies slowly as a function of the 
one neutron separation energy and becomes smaller than 
that of the $^{31}$Ne nucleus for $\epsilon_n \leq -0.42$ MeV due to the 
pairing anti-halo effect. 
This behavior is qualitatively the same as in the previous analysis 
shown in the middle panel of Fig. 2 in Ref. \cite{HS11}, 
which was based on the spherical symmetry of the Ne isotopes. 

The reaction cross sections for the 
$^{30,31,32}$Ne nuclei evaluated at 
$S_n (^{31}$Ne) = 0.3 MeV 
are shown in Fig. 2. 
These are compared with the experimental interaction 
cross sections \cite{Takechi12}, which are expected to be close to the 
reaction cross sections for neutron-rich nuclei 
\cite{Takechi14,OYS92,Kohama08}. 
For comparison, we also show the result of the previous analysis \cite{HS11} 
at a similar one neutron separation energy, even though the single-potential 
is different from the one used in the present analysis. 
One can see that the odd-even staggering can still be reproduced 
by taking into account the nuclear deformation. Notice that the degree of 
the staggering becomes smaller in the present calculation 
compared to the previous spherical calculation. 
This is because the valence neutron in $^{31}$Ne fully occupies the 1$p_{3/2}$ 
level in the spherical case, while the occupation probability for 
the $p_{3/2}$ level decreases from unity in the deformed case. 
In the case shown in Fig. 2, 
our calculation yields the occupation probability of 46.3 \%. 

\begin{figure} 
\includegraphics[scale=0.6,clip]{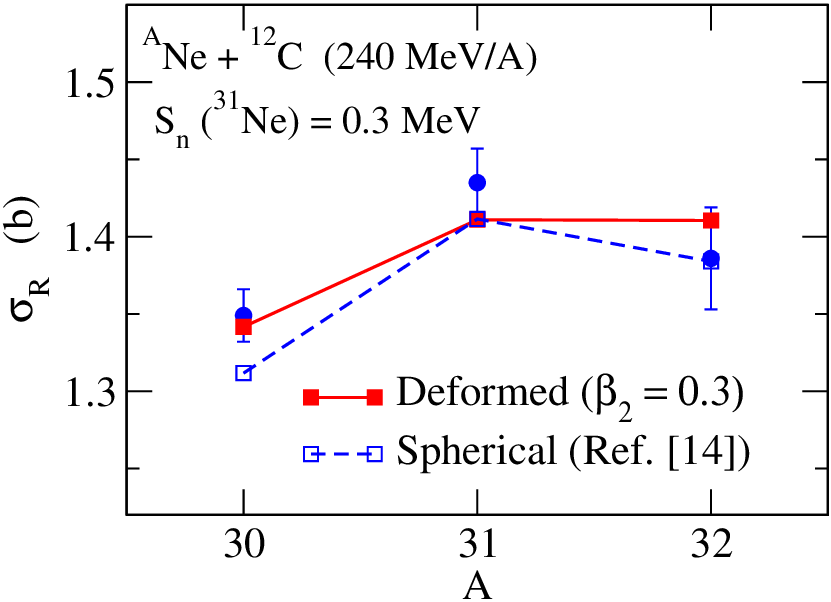}
\caption{The reaction cross sections for the 
$^{30,31,32}$Ne + $^{12}$C reaction at $E$ = 240 MeV/nucleon. 
These are evaluated at the one neutron separation energy of 
$^{31}$Ne of $S_n$ = 0.3 MeV. 
The filled circles with error bars indicate the experimental 
interaction cross sections taken from Ref. \cite{Takechi12}. 
For comparison, the result of the previous analysis \cite{HS11} 
based on the spherical 
density distributions 
is also shown by the dashed line. } 
\end{figure}

A larger staggering can be obtained with deformation 
when the one neutron separation energy 
of $^{31}$Ne is further decreased. 
In order to demonstrate this, Fig. 3 shows the 
staggering parameter $\gamma_3$ defined as \cite{HS12},
\begin{equation}
\gamma_3\equiv 
\frac{(-1)^A}{2}\,[\sigma_R(A+1)-2\sigma_R(A)+\sigma_R(A-1)],
\label{gamma3}
\end{equation}
where $\sigma_R(A)$ is the reaction cross section of a nucleus with 
mass number $A$. 
The dashed and the solid lines in the figure 
show the staggering parameter for 
the spherical and the deformed cases, 
respectively. 
For a fixed value of separation energy, 
the staggering parameter $\gamma_3$ is smaller in the deformed case as 
compared to the spherical case, which is consistent with Fig. 2. 
However, the staggering parameter 
increases as the separation energy decreases, and 
eventually comes closer to the 
central value of the experimental data when $\epsilon_n$ is around the 
threshold. 

\begin{figure} 
\includegraphics[scale=0.6,clip]{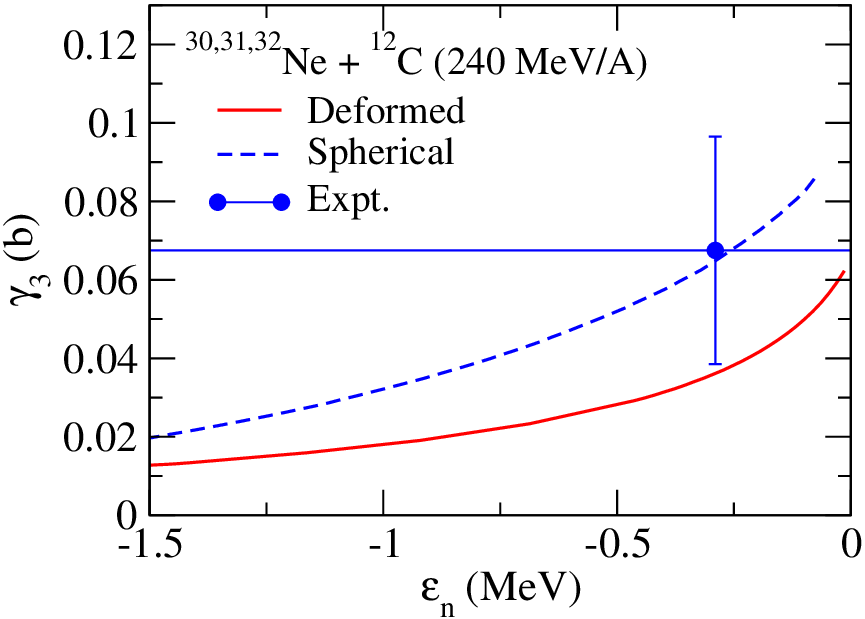}
\caption{The staggering parameter $\gamma_3$ defined by Eq. (\ref{gamma3}) 
for the $^{30,31,32}$Ne + $^{12}$C reactions at $E$ = 240 MeV/nucleon. 
It is plotted as a function of the single-particle energy for the 
valence orbit for the $^{31}$Ne nucleus. The dashed and the solid lines 
are obtained using the spherical and the deformed density distributions, 
respectively. The experimental data is evaluated using the measured 
interaction cross sections shown in Ref. \cite{Takechi12}. 
}
\end{figure}

In Sec. I, we have conjectured that the nuclear deformation may lead to 
two effects on reaction cross sections. One is to decrease 
the cross section for $^{31}$Ne due to the admixture of several angular 
momentum components in the single-particle wave function for the valence 
orbit. 
The other is to change the cross section for $^{32}$Ne 
because of a change in the degree of pairing anti-halo effect. 
The former effect 
would reduce the staggering, while the latter effect either enhances or 
reduces it depending on the level density around the Fermi surface. 
Our calculation shown in Fig. 2 indicates that the former effect indeed 
exists, while the latter effect is much less clear. 
In order to shed light on the latter effect, 
Fig. 4 shows the dependence of the reaction cross section on the 
strength of the pairing correlation. To this end, we vary the average 
pairing gap, $\bar{\Delta}$, defined by Eq. (\ref{avdelta}). 
The filled squares with the dashed, the solid, and the dotted lines 
are obtained by setting the average pairing gap to be 1, 
2.1 (=12/$\sqrt{32}$), and 3 MeV, 
respectively. Notice that the solid line is the same as that in Fig. 2. 
For comparison, the open square 
with the dashed line shows the result 
without the pairing correlation. 
One can clearly see that the reaction cross section for $^{32}$Ne 
is not sensitive to the value of average pairing gap 
as long as it is large enough. 
This would be 
correlated to the fact that the root-mean-square radius for $^{32}$Ne 
is not sensitive to the single-particle energy for the valence orbit of 
$^{31}$Ne, as has been shown in Fig. 1. 
Even though the occupation probability for the $p$-wave orbital may depend 
largely on the average pairing gap, 
the root-mean-square radius does not change much once the radius is 
significantly shrunken due to the pairing anti-halo effect so that 
the $s$ and $p$-wave states do not behave abnormally. 
This indicates that the main effect of nuclear deformation is simply 
to decrease 
the odd-even staggering in reaction cross sections, at least 
for the Ne isotopes. 

\begin{figure} 
\includegraphics[scale=0.6,clip]{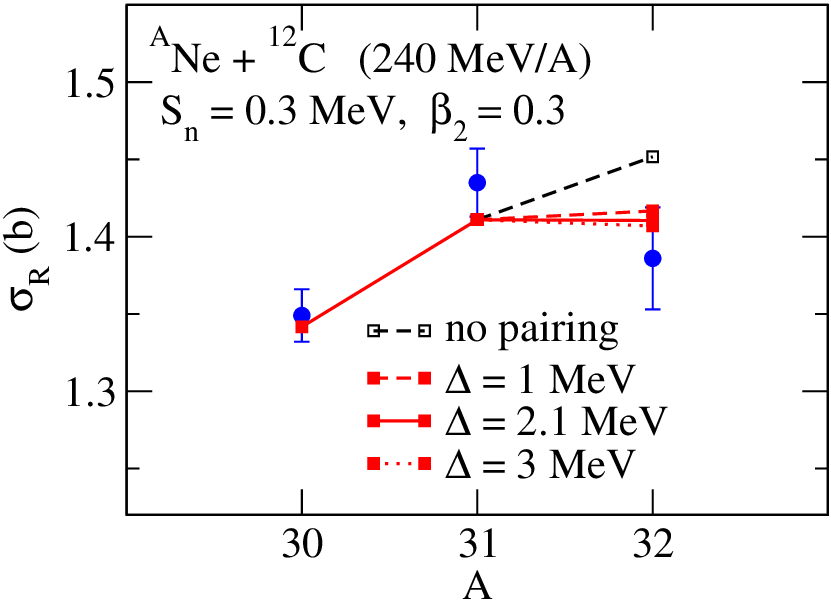}
\caption{The dependence of the reaction cross section for the 
$^{32}$Ne nucleus on the average pairing gap. 
The filled squares with the dashed, the solid, and the dotted lines 
are obtained by setting the average pairing gap to be 1, 2.1 (=12/$\sqrt{32}$), 
and 3 MeV, 
respectively. For comparison, the open square with the dashed line 
shows the result without the pairing correlation. 
}
\end{figure}

\subsection{$^{36,37,38}$Mg isotopes}

Let us next discuss the $^{36,37,38}$Mg isotopes. 
For $^{37}$Mg in these isotopes, the $p$-wave halo structure has 
been suggested from a measurement of the one neutron removal 
reaction on C and Pb targets, with a small one neutron separation 
energy of 0.22$^{+0.12}_{-0.09}$ MeV \cite{Kobayashi14}. 
Moreover, the experimental reaction cross sections indicate a large 
odd-even staggering for $^{36,37,38}$Mg \cite{Takechi14}. 

We first determine the deformation parameter $\beta_2$ 
for these isotopes. 
With the deformed Woods-Saxon potential shown in Sec. II-A, together 
with the parameters listed in Table I in Ref. \cite{Shoji09}, 
we find that the valence neutron in $^{37}$Mg 
occupies the [312 5/2] orbit for $\beta_2 \leq$ 0.4, while it occupies 
the [321 1/2] orbit for $0.4 \leq \beta_2 \leq 0.6$. 
The former is connected to the 1$f_{7/2}$ level while the latter to the 
2$p_{3/2}$ level in the spherical limit. The former state has $K^\pi=5/2^-$, 
and thus contains angular momenta larger than $l=3$, which do not form a 
halo structure. In contrast, the latter state contains a large 
$p$-wave component, being consistent with the suggested halo structure for 
$^{37}$Mg \cite{Takechi14}. We therefore choose $\beta_2=0.5$ in 
the analysis shown below. 

Figure 5 shows the root-mean-square radii for the 
$^{36,37,38}$Mg nuclei as a function of the 
single-particle energy for the valence orbit of $^{37}$Mg. 
The radii behave qualitatively the same as those for 
the Ne isotopes shown in Fig. 1. That is, 
the radius of $^{37}$Mg diverges in the limit of vanishing 
single-particle energy, while that of $^{38}$Mg 
varies much more slowly due to the pairing anti-halo effect. 

\begin{figure} 
\includegraphics[scale=0.6,clip]{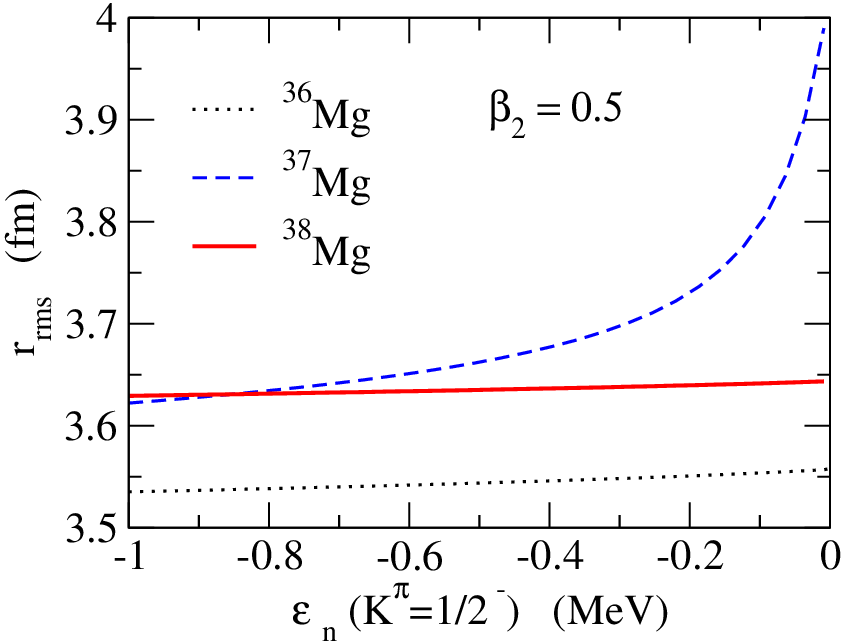}
\caption{
Same as Fig. 1, but for the 
$^{36,37,38}$Mg isotopes. 
}
\end{figure}

The reaction cross sections for the 
$^{36,37,38}$Mg isotopes are shown in Fig. 6 for two different values of the 
one neutron separation energy, $S_n$, for $^{37}$Mg. 
The solid line is obtained with $S_n$ = 0.32 MeV, while the dashed line 
with $S_n$ = 1.5 MeV. The experimental odd-even staggering can be well 
reproduced with 
$S_n$ = 0.32 MeV. On the other hand, 
for $S_n$ = 1.52 MeV, 
the reaction cross section increases 
monotonically as a function of mass number, 
that is 
inconsistent with the experimental data. 
Notice that this behavior is qualitatively similar to the odd-even staggering for 
the Ne isotopes 
shown in Fig.3 in Ref. \cite{HS11} obtained with the spherical densities. 
Therefore it is evident that 
the pairing anti-halo effect plays an important role in 
the odd-even staggering of deformed neutron-rich nuclei, such as 
$^{30,31,32}$Ne and $^{36,37,38}$Mg isotopes. 

\begin{figure} 
\includegraphics[scale=0.6,clip]{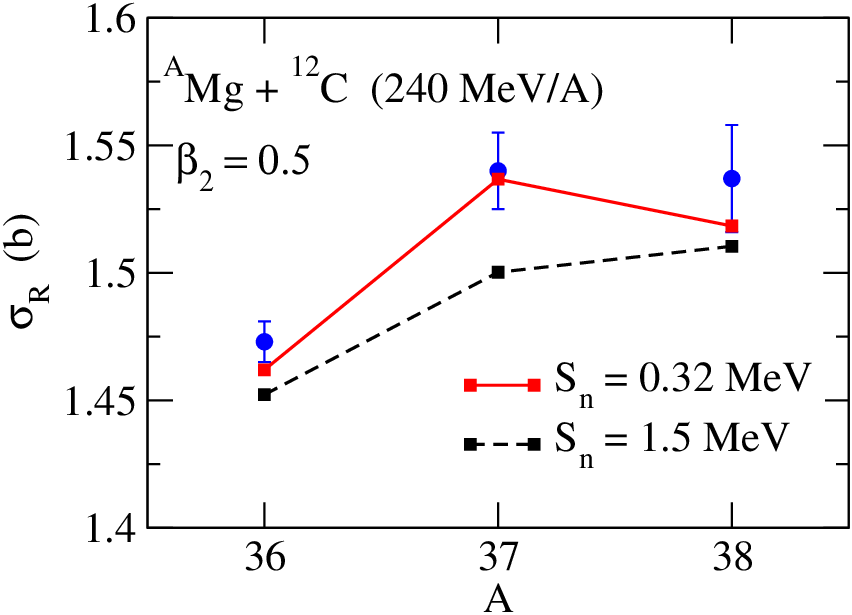}
\caption{The reaction cross sections for the 
$^{36,37,38}$Mg + $^{12}$C reaction at $E$ = 240 MeV/nucleon. 
These are evaluated 
for the quadrupole deformation parameter of $\beta_2$ = 0.5, 
at which the valence neutron in $^{37}$Mg occupies the 
[321,1/2] orbital. 
The solid and the dashed lines denote the results for 
$S_n$ = 0.32 and 1.5 MeV, respectively, where 
$S_n$ is the one neutron separation energy of 
$^{37}$Mg. 
The experimental reaction cross sections are 
taken from Ref. \cite{Takechi14}. 
}
\end{figure}

\section{Summary}

We have investigated the role of nuclear deformation in the 
odd-even staggering observed in reaction cross sections for several systems. 
To this end, we have used the deformed Hartree-Fock-Bogoliubov method 
to take into account both the deformation and the pairing effects. 
We have applied this method to the 
$^{30,31,32}$Ne and $^{36,37,38}$Mg isotopes 
and have shown that 
the deformation mainly decreases the degree of odd-even staggering due 
to the admixture of several angular momentum states in a deformed 
single-particle wave function. Despite this, the odd-even staggering persists 
even with finite deformation, when the one neutron separation energy is small 
enough. In particular, we have successfully accounted for the experimental 
odd-even 
staggering both for the 
$^{30,31,32}$Ne and $^{36,37,38}$Mg isotopes within the unified theoretical 
framework. 
This strongly indicates that the pairing anti-halo effect 
indeed has a responsibility 
to the observed odd-even staggering in reaction cross sections. 

Our calculation can be improved in several ways. For instance, in this 
paper, we have assumed that the deformation 
parameter is the same for the three 
nuclei within the same element. It might be important to take into account 
an isotope dependent deformation, as has been predicted e.g., 
by the anti-symmetrized molecular dynamics (AMD) 
\cite{Minomo11,Minomo12,Minomo12-2,Watanabe14}, 
although the 
dependence of the reaction cross sections on the deformation 
would not be large once the 
single-particle configuration is fixed. 
Another issue is the treatment of pairing for the odd-mass nuclei. 
For simplicity, in this paper we have neglected the pairing correlation 
in $^{30,31}$Ne and $^{36,37}$Mg, because the effect of the pairing on the 
radius of $^{30}$Ne and $^{36}$Mg had turned out to be small. 
However, if one regards 
$^{31}$Ne and $^{37}$Mg as one quasi-particle excitation on top of 
$^{32}$Ne and $^{38}$Mg, the pairing might play some role in these nuclei 
as well. 
A more consistent way towards this end 
would be to treat the odd-mass nuclei using the blocked 
HFB method \cite{Bally14,Matsuo13}, that would be an interesting future work. 

\section*{Acknowledgments}
This work was partly supported by JSPS KAKENHI Grant Numbers 16H02179 
and JP16K05367.

\end{document}